# Acoustic and optical phonon dynamics from femtosecond time-resolved optical spectroscopy of superconducting iron pnictide $Ca(Fe_{0.944}Co_{0.056})_2As_2$

Sunil Kumar,[1] L. Harnagea,[2] S. Wurmehl,[2] B. Buchner[2] and A. K. Sood[1,*]

[1]*Department of Physics and Center for Ultrafast Laser Applications, Indian Institute of Science Bangalore 560012, India*
[2]*Leibniz-Institute for Solid State and materials Research, Dresden, D-01171 Dresden, Germany*

We report temperature evolution of coherently excited acoustic and optical phonon dynamics in superconducting iron pnictide single crystal $Ca(Fe_{0.944}Co_{0.056})_2As_2$ across the spin density wave transition at $T_{SDW} \sim 85$ K and superconducting transition at $T_{SC} \sim 20$ K. Strain pulse propagation model applied to the generation of the acoustic phonons yields the temperature dependence of the optical constants, and longitudinal and transverse sound velocities in the temperature range of 3.1 K to 300 K. The frequency and dephasing times of the phonons show anomalous temperature dependence below $T_{SC}$ indicating a coupling of these low energy excitations with the Cooper-pair quasiparticles. A maximum in the amplitude of the acoustic modes at $T \sim 170$ is seen, attributed to spin fluctuations and strong spin-lattice coupling before $T_{SDW}$.

**Introduction.** − Recent discovery of high-temperature superconductivity in iron pnictides [1,2,3] is being studied extensively to understand the quasiparticle pairing in these compounds in comparison with the cuprates. Various experimental techniques such as infrared optical spectroscopy [4,5,6] and angle resolved photo-emission spectroscopy [7,8] have revealed multiple charge gaps in the superconducting phase. The response of iron pnictides to ultrafast optical excitation can contribute to the knowledge of the superconductivity in these systems. In the last couple of years, much of the time-resolved optical studies on iron pnictides have focused on the carrier dynamics providing valuable information on the relaxation pathways, the multi-band structure and electron-optical phonon coupling in these systems [9,10,11,12,13]. However, very few studies are available on the coherent phonons in these systems [9,14,15] where knowledge of the temperature dependence of these low-energy excitations around the superconducting transition temperature will be essential to understand their role in superconductivity.

In this letter we present experimental results on temperature evolution of femtosecond time-resolved differential reflectivity of superconducting $Ca(Fe_{0.944}Co_{0.056})_2As_2$ single crystals from the Ca-122 group of iron pnictide systems not explored hitherto. The transient reflectivity signals clearly show fast (sub-ps time-period) and slow (~100 ps time-period) coherent oscillations along with three-component electronic relaxation with time-constants varying from sup-ps to few ns. Fast coherent oscillations with a frequency of ~5.7 THz are attributed to the $A_{1g}$-symmetric coherent optical phonon mode in the crystal. For the first time, in the complete temperature range of 3.1 to 300 K the transient reflectivity signals have revealed long wavelength coherent oscillations superimposed on slowly exponentially decaying background. These are attributed to longitudinal and transverse acoustic phonons in the system. The present study is the first report on investigations of coherent optical and acoustic phonon dynamics in all three phases: the high-temperature normal metal (T > 85 K), the spin density wave (SDW) phase at the intermediate temperatures (85 K to 20 K) and superconducting (SC) phase at temperatures lower than 20 K. Our observations of these GHz frequency acoustic phonons discussed in the light of propagating strain pulse model [16] yield the temperature-dependence of longitudinal and transverse sound velocities which can be used to estimate the elastic moduli along by knowing the temperature dependence of the material density. Large variations in the phonon parameters around the SDW and SC phase transition temperatures indicate strong magneto-elastic interaction.

**Experimental details.** – Single crystals of $Ca(Fe_{1-x}Co_x)_2As_2$ with x = 0.056 used in the present study were grown by high temperature solution growth technique using Sn flux and cleaved into platelet samples with thickness ~0.5 mm and c-axis perpendicular to the surface. A complete characterization for the crystal has been reported earlier [17]. The superconducting and SDW phase transitions occur at temperatures $T_{SC} \sim 20$ K and $T_{SDW} \sim 85$ K. The crystal shows concurrent structural transition from the high temperature tetragonal to low temperature orthorhombic phase at $T_S \sim 88$ K. Degenerate pump-probe experiments in a noncollinear geometry were carried out at 790 nm using 45 fs laser pulses taken from 1 kHz repetition rate regenerative amplifier. The pump and probe polarizations were kept orthogonal to each other and an analyzer orthogonal to the pump polarization was placed at the detector. Our pump and probe beam diameters are quite large ~1.9 mm (measured using a beam profiler placed at the sample point) covering about 80% of the sample surface area. Transient differential reflectivity signals were systematically recorded as a function of time-delay between the pump and the probe pulses at various pump-fluences varying from ~25 μJ/cm$^2$ to 220 μJ/cm$^2$ and sample temperatures from 3.1 K to 300 K with the sample mounted inside a continuous flow liquid-helium optical cryostat. During the entire set of experiments, the pump-probe beam polarizations and the angle of incidence as well as the orientation of the crystal inside the cryostat were not changed. The reproducibility of the experimental data has been ensured by repeating the experiments. We performed additional linear reflectivity measurements by focusing the femtosecond laser beam (low average power) on the flat portion of the sample and collecting all the reflected light for near-normal incidence. The measured value of the reflectivity R is 0.32 which is constant as a function of temperature within an accuracy of ±5%.

**Results and discussion.** – Experimental time-resolved differential reflectivity data at various sample-temperatures taken at a moderate value of the pump-fluence of ~80 μJ/cm$^2$ are presented in Fig. 1 where the temperature dependence of various exponentially decaying and oscillatory components can be clearly seen at faster time-scale (left panel) and longer time-scale (right panel). Overall, the transients can be decomposed into an electronic part (exponentially decaying) and a coherent phononic (damped oscillatory) part fitted using the following function convoluted with the Gaussian laser pulse,

$$\frac{\Delta R}{R} = \left(1 - e^{-t/\tau_r}\right)\left[\sum_k A_k e^{-t/\tau_k} + \sum_l B_l e^{-t/\tau^p_l} \cos(\omega_l t + \phi_l)\right] \quad (1)$$

The term inside the small brackets takes care of the initial rise of the signals with rise time of $\tau_r$ ~ 80 fs, found to be the same for all the measurements. The data can be consistently fitted (see supplementary material) with three electronic components ($k$ = 1,2,3) with amplitude $A_k$ and time-constant $\tau_k$, and a combination of one fast oscillatory component ($l$ = 1) attributed to coherent optical phonon mode and two slow oscillatory components ($l$ = 2,3) attributed to longitudinal and transverse acoustic phonons. The parameters $B_l$, $\tau^p_l$, $\omega_l$ and $\phi_l$ represent the amplitude, dephasing time, angular frequency and the initial phase of the phonon oscillations.

In the present paper we have focused our discussion on the temperature evolution of the fast (sub-ps time period) and slow (~100 ps time-period) oscillatory modes in the experimentally obtained transient differential reflectivity signals. The carrier dynamics will be reported elsewhere [18]. The temperature dependence of the two long-wavelength oscillatory modes attributed to the strain pulse generated longitudinal and transverse acoustic phonon modes has been discussed below followed by the results on the coherent optical phonon mode centered at a frequency of ~5.7 THz at 3.1 K. The temperature dependent frequency and dephasing time of the coherent longitudinal acoustic mode have been fully analyzed in the context of strain pulse propagation model, which, in conjunction with separately measured reflectivity value R yields the material refractive index n and extinction coefficient κ (the real and imaginary parts of the complex index n$^*$ = n + iκ), and hence the optical penetration depth ζ at 790 nm as a function of temperature. We estimate ζ to be ~40 nm at the room temperature which increases to a value of ~200 nm at 3.1 K. We note that the room temperature value of ζ in other 122-type pnictides is reported to be between 25 to 60 nm [12,13,19].

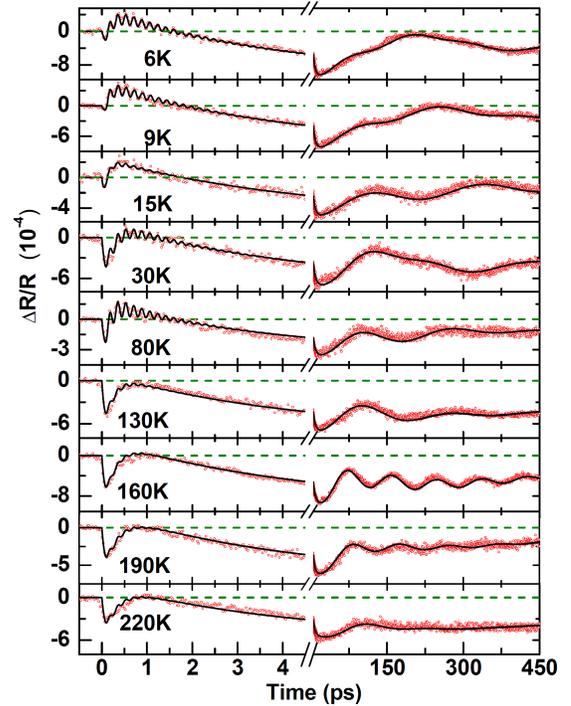

**Fig.1.** (Color online) Transient differential reflectivity spectra of Ca(Fe$_{0.944}$Co$_{0.056}$)$_2$As$_2$ at various sample temperatures. Solid lines are fits using Eq. 1.

**Longitudinal and transverse acoustic phonons.** – Figure 2 presents the temperature-dependence of the amplitude $B_l$, dephasing time $\tau^p_l$, frequency $\nu_l$ and initial phase $\varphi_l$ of the two long-wavelength coherent oscillations corresponding to $l$ = 2,3 in Eq. 1. Large temperature-dependence of each parameter is evident from Fig. 2, however, they are found to be fluence-independent except the amplitude $B_l$ which linearly scales with the pump-fluence. This particular oscillatory contribution in the transient reflectivity arises due to propagating acoustic waves in the crystal generated by laser induced transient stress, either thermal or electronic [16]. Thus produced strain-pulse can propagate through the entire thickness of the crystal; however, its propagation is detected only within the optical penetration depth at the probe wavelength. Previously, highly damped coherent longitudinal acoustic phonons with frequency ~25 GHz have been reported only in the superconducting hole-doped (K-doped) Ba-122 systems where the amplitude of this mode was found to be much smaller in the superconducting state due to sharp decrease in the thermal expansion coefficient at $T_{SC}$ [12], whereas, in the electron-doped (Co-doped) Ba-122 systems, this mode was not observed which was believed to be due to suppressed thermal expansion coefficient along the c-axis [12]. This is in contrast to our experimental results where both the undoped [20] and the doped (present study) compounds from the Ca-122 family clearly show coherent acoustic phonons at all temperatures having the largest amplitude at a temperature higher than $T_{SDW}$. This indicates that the thermal expansivity of the Ca-122 systems as a function of temperature may be different from that of the Ba-122 systems.

The higher frequency mode in Fig. 2 is attributed to the longitudinal acoustic LA and the lower one to the shear or transverse acoustic TA. At any temperature T, the frequencies are related to the corresponding sound velocities ($u_{TA,LA}$) in the crystal through the relation [21]

$$\nu_{TA,LA}(T) = \frac{2n(T)u_{TA,LA}(T)\cos\theta}{\lambda_{pr}}, \quad (2)$$

n being the refractive index and θ the angle of incidence of the probe beam of central wavelength $\lambda_{pr}$. Knowing refractive index n, the frequency of the observed acoustic phonon modes provides an estimation of the corresponding sound velocities $u_{TA,LA}$.

Excitation of both the LA and TA modes in a single crystal by the strain pulse propagation mechanism has been seen in many studies [22,23,24,25]. For example, in a recent study on multiferroic BaFeO$_3$ single crystal, the Brillouin frequencies of the detected LA and TA phonon components have been used to estimate the one longitudinal and two transverse sound velocities [22]. In the strain pulse model, the amplitude of the phonon oscillation depends only on pump-photon-energy and pump-fluence whereas the frequency, dephasing time and phase of the oscillations depend on the angle of incidence and the wavelength of the probe $\lambda_{pr}$ [21,22,26]. While the initiation of the longitudinal motion is not unusual, thermoelastic excitation of the shear motion requires crystals asymmetrically cut with symmetry axes away from the surface normal [23,24]. A small miscut is sufficient to induce shear motion along the thickness of a single crystal [25]. Our experiments on undoped and Co-doped Ca-122 compounds for doping levels of x = 0.056 (present study) and 0.073 (using pump pulses centered at 400 nm and probe pulses centered at 790 nm) show that a single mode attributed to the longitudinal acoustic phonons, is observed in the undoped CaFe$_2$As$_2$ compound [20], however, two modes, one longitudinal and one transverse, are observed in the doped compounds. We do not have a quantitative understanding of this difference yet.

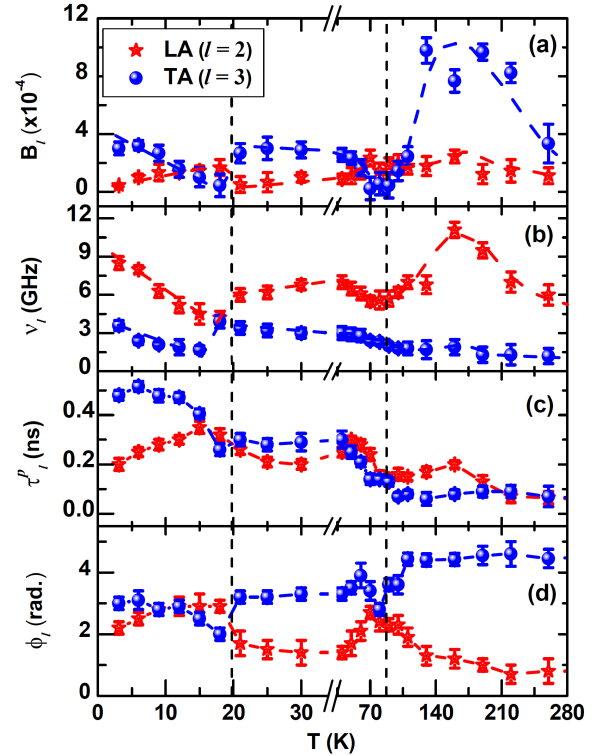

**Fig.2.** Temperature-dependence of the amplitude $B_l$, damping time $\tau^p_l$, frequency $\nu_l$ and initial phase $\phi_l$ of the experimentally observed longitudinal acoustic (LA) and transverse acoustic (TA) modes corresponding to $l$ = 2, 3 in Eq. 1. The dashed lines in each panel are guide to the eyes and the dotted vertical lines mark the nominal superconducting and SDW transition temperatures.

We notice from Fig. 2(a) that in the normal state (T > 90 K), the amplitudes of both the LA and TA modes are higher than those at lower temperatures (T < $T_{SDW}$), particularly a maximum in $B_{TA}$ at a temperature ~170 K. Such a feature can arise due to precursors of spin fluctuations and strong spin-lattice coupling before the SDW state is setup at $T_{SDW}$ [27,28,29]. Similar precursor effect has been attributed to play a role in enhancing the amplitude of coherent longitudinal acoustic phonons above the magnetic transition temperature in multiferroic manganites [30,31]. Furthermore, competing nature of the amplitudes of the two phonon modes, i.e., the amplitude of the LA mode being maximum when the amplitude of the TA mode is minimum and vice-versa, can also be seen from Fig. 2a below $T_{SDW}$. This result needs to be understood better by taking into consideration the changes in the interactions among electronic, phonon and spin degrees of freedom while going from the normal paramagnetic-metallic state to the superconducting state through an intermediate SDW state.

The phonon-phase $\varphi_l$ of each of the two modes shows abrupt changes at $T_{SC}$ and $T_{SDW}$ (Fig. 2d). For our sample the structural transition from high-temperature tetragonal to low temperature orthorhombic phase occurs at $T_S \sim 88$ K which is very close to $T_{SDW} \sim 85$ K. Magneto-elastic coupling induced suppression of orthorhombic distortions in the superconducting phase is known for the doped iron pnictides [32]. Therefore, the observed change in the phonon-phase around $T_{SDW}$ and $T_{SC}$ is a signature of the underlying changes in the crystalline structure and magneto-elastic coupling in the Ca-122 lattice at the SDW and superconducting phase transitions.

The frequency $\nu_l$ and the dephasing time $\tau^p_l$ of the two modes are also highly temperature dependent and show drastic changes at the $T_{SC}$ and $T_{SDW}$ (Figs. 2b and 2c). The dephasing time of the acoustic phonons can arise mainly from two factors [26]; the intrinsic life time $\tau_{phonon}$ and damping $\tau_\alpha$ due to probe absorption coefficient α within the optical penetration depth $\zeta$ so that $1/\tau = 1/\tau_{phonon} + 1/\tau_\alpha$ where $\tau_\alpha = \zeta/u$, $u$ being the sound velocity. In thick crystals (as in our case) the observed phonon dephasing of the LA mode should be related to the probe penetration depth along the thickness in the crystal. Therefore, identifying $\tau_{LA}$ in Fig. 2c as $\tau_\alpha$, in the context of the strain pulse propagation model, at any temperature T the penetration depth $\zeta$ at the probe wavelength $\lambda_{pr}$ is related to $\tau_{LA}$ and the corresponding sound velocity $u_{LA}$ via the relation $\zeta(T) = u_{LA}(T)\tau_{LA}(T)$. Since $\zeta = \lambda_{pr}/4\pi\kappa$, where κ is the extinction coefficient, this relation for $\zeta$ and Eq. 2 (for θ = 0) simply lead to $n(T)/\kappa(T) = 2\pi\nu_{LA}(T)\tau_{LA}(T)$. For normal incidence, the Fresnel's relation for reflectivity is given as $R = \{(n-1)^2 + \kappa^2\}/\{(n+1)^2 + \kappa^2\}$. Using constant value of R = 0.32±0.02 during 3.1 K to 300 K, we have estimated κ and n as a function of T from the above analysis and are shown in Figs. 3a and 3b. These have been used to calculate the temperature dependent longitudinal sound velocity $u_{LA}$ and optical penetration depth $\zeta$ as presented in Figs. 3c and 3d, respectively.

The result for the temperature dependent refractive index n(T) in Fig. 3b and the frequency of the transverse mode $\nu_{TA}$ in Fig. 2b can be used to calculate the transverse sound velocity $u_{TA}$ using Eq. 2 which has the same temperature dependence as $\nu_{TA}$ at temperatures below $T_{SDW}$ where n(T) is nearly temperature independent. Knowing the material density ρ, the elastic modulus along the longitudinal and transverse directions can be calculated via relation $Y_s = \rho u_s^2$ where index s = LA or TA. The temperature-dependent material density in doped Ca-122 compounds has yet to be studied. For the undoped $CaFe_2As_2$ compound, a sharp increase in the volume of the lattice unit cell (hence decrease in the density) was observed at $T_{SDW}$ while increasing the temperature [33]. Assuming the same trend for the doped Ca-122 system, we see that along the longitudinal direction the elastic softening at $T_{SDW}$ will be even more. The elastic softening around $T_{SC}$ and $T_{SDW}$ has been seen previously in Ca-122 compounds using ultrasound experiments at kHz to MHz frequencies [34,35], however, the difference in the values of the elastic modulus from the two experiments (the present ultrafast ultrasonics measurements at GHz vis-à-vis previous ultrasound experiments at kHz and MHz) could be possibly arising from the difference in the coupling of sound waves of two different frequencies with the slow relaxation processes in the system [36].

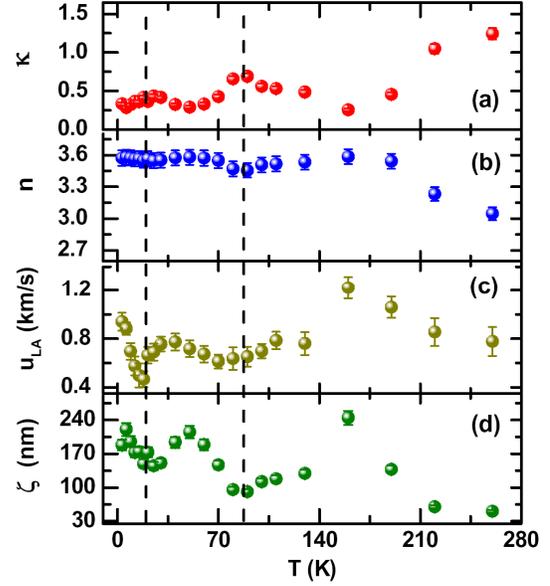

**Fig.3.** Temperature-dependent (a) extinction coefficient, (b) refractive index, (c) longitudinal sound velocity, and (d) optical penetration depth at 790 nm estimated from the stain pulse propagation model applied to the experimentally observed LA mode vis-à-vis the temperature independent reflectivity value R as described in the text.

Elastic softening is a generic feature close to a structural transition and is indicative of strong magneto-elastic coupling [35,37]. In our crystal, the tetragonal to orthorhombic phase transition occurs at ~88 K around which we have observed elastic softening in the LA branch. However, the elastic anomalies below $T_{SC}$ (Fig. 2b) are very unusual which can arise only due to structural instability in the lattice [35]. The question to be asked is whether orbital fluctuations [38] can be invoked to explain the observed elastic anomalies in superconducting Ca-122 pnictides.

**Coherent optical phonons.** – Now we turn to the fast oscillations in Fig. 1 ($l = 1$ in Eq. 1) which are attributed to the $A_{1g}$ symmetric coherent optical phonon mode similar to that reported in other 1111 (Refs. [9,14]) and 122 (Ref. [15]) type of pnictides. There are four Raman-active phonon modes in the 122-type compounds with irreducible representations $A_{1g}$, $B_{1g}$ and $2E_g$ [39]. The observed mode in our experiments at 5.7 THz is attributed to the $A_{1g}$-symmetric optical phonon mode having only breathing displacements of the As atoms.

The displacive excitation of the coherent phonons (DECP) mechanism can be invoked for the coherent phonon generation [15].

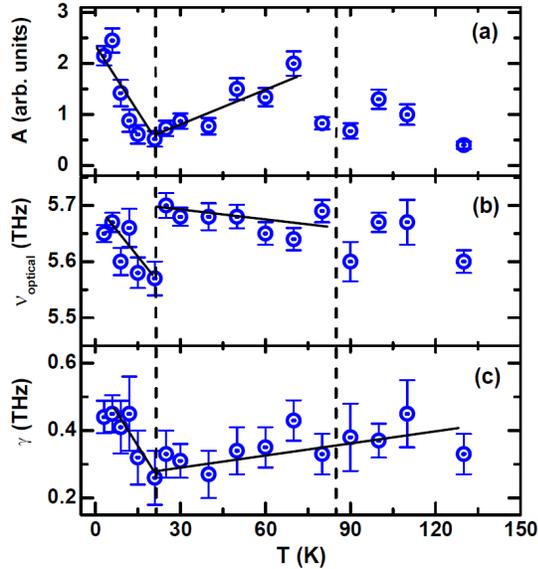

**Fig.4.** Single Lorentzian fit parameters, the amplitude A, frequency $\nu_{optical}$ and spectral broadening $\gamma$ for the data obtained by fast Fourier transform of coherent phonon oscillations in the Fig. 1. Solid lines are drawn as guide to the eyes.

Single Lorentzian-oscillator fitting to fast Fourier transform of the differentiated raw data, taken up to 3 ps in Fig. 1, gives the temperature evolution of the optical phonon amplitude A, frequency $\nu_{optical}$ and life time broadening $\gamma$ as presented in Fig. 4. After a short laser pulse excitation, the photogenerated carrier density relaxes exponentially so that the force function for the generation of coherent phonons can be assumed simply an exponential decay function with time constant equal to the fastest electronic relaxation time. Comparing the fastest electronic decay time $\tau_1$ ~150 fs in the superconducting state of our system with the phonon dephasing time $1/\pi\gamma$ from Fig. 4(c), we find that $\tau_1$ is much smaller than $1/\pi\gamma$, hence ensuring coherent energy transfer [40] from fast electrons to the optical phonon mode in the iron pnictide crystal. As the temperature approaches $T_{SC}$ from below, we find that $\tau_1$ slows down from ~100 fs to ~200 fs, thereby decreasing the production of the optical phonons as reflected from the decreasing amplitude A in Fig. 4a. The frequency $\nu_{optical}$ and the line-width $\gamma$ of the observed optical mode decrease anomalously with the increasing temperature as T approaches $T_{SC}$ followed by a discontinuous jump in the frequency at $T_{SC}$ (Fig. 4b), a behavior similar to that observed for the frequency of the two acoustic modes discussed before in Fig. 2b. This particular observation suggests that the superconducting quasiparticles are coupled to the optical and possibly with the acoustic phonons also, in the Ca-122 lattice.

Previously, the $A_{1g}$-symmetric coherent optical phonon in superconducting Ba-122 compound were observed using very high pump-fluences (~1 to 6 mJ/cm$^2$) where no changes were observed across the superconducting transition temperature [15]. This can be due to the fact that high pump-fluences drive the system into the normal state so that the observed phonon is insensitive to Cooper-pair recombination dynamics. On the other hand, we clearly see that using a moderate pump-fluence of ~80 μJ/cm$^2$, the $A_{1g}$-symmetric optical phonon and the two acoustic phonon parameters show large changes across the superconducting transition temperature in underdoped Ca-122 compound indicating a coupling of the Cooper-pair quasiparticles with phonons.

An interaction between the phonons and superconducting charged quasiparticles leads to change in the phonon self-energy [41]. The anomalous softening and a discontinuous jump in the frequency of the observed $A_{1g}$ phonon at $T_{SC}$ (Fig. 4b) can be attributed to coupling of this phonon with the superconducting quasiparticles. In that case, the superconducting gap value $2\Delta_0$ can be approximated with $\nu_{optical}$ resulting in $2\Delta_0/k_B T_{SC}$ ~ 13.5, a value much larger than the BCS-mean value. We note that the large variation in the $2\Delta_0/k_B T_{SC}$ value from 1 to 10 from different experiments is still not understood though multiple charge gap opening at $T_{SC}$ has been consistently observed [4,5,6,7,8].

**Conclusions.** – In conclusion, transient differential reflectivity of superconducting iron pnictide $Ca(Fe_{0.944}Co_{0.056})_2As_2$ was studied to extract the dynamics of coherent optical phonons along with the longitudinal and transverse acoustic phonons in the crystal. Strain pulse propagation model applied to the experimentally observed longitudinal acoustic mode yields the temperature-dependence of optical constants and the sound velocity as well as elastic behavior of the crystal in the temperature range of 3.1 to 300 K. Significant changes in the phonon parameters at $T_{SC}$ and $T_{SDW}$ indicate a coupling of these low-energy material excitations with charge quasiparticles which will contribute to the understanding of their role in the origin of superconductivity from the high-temperature spin ordered phase in these systems.

AKS and SK acknowledge Department of Science and Technology, India for financial assistance. The materials were fabricated and characterized at "Leibniz-Institute for Solid State and materials Research" supported by the Deutsche Forschungsgemeinschaft through the Priority Programme SPP1458 (Grant No. BE1749/13), and through the Emmy Noether Programme WU595/3-1 (S.W.).

# Acoustic and optical phonon dynamics from femtosecond time-resolved optical spectroscopy of superconducting iron pnictide Ca(Fe$_{0.944}$Co$_{0.056}$)$_2$As$_2$ : Supplementary material

SUNIL KUMAR, L. HARNAGEA, S. WURMEHL, B. BUCHNER and A. K. SOOD

To fit our time-resolved differential reflectivity data, we have used the following function convoluted with the Gaussian laser pulse,

$$\frac{\Delta R}{R} = \left(1 - e^{-t/\tau_r}\right)\left[\sum_k A_k e^{-t/\tau_k} + \sum_l B_l e^{-t/\tau^p_l} \cos(\omega_l t + \phi_l)\right] \quad (1)$$

The data can be consistently fitted with three electronic components ($k = 1,2,3$) and three oscillatory components ($l = 1,2,3$). The fitting procedure is shown in the Fig. S1(a) below for sample temperature 6 K where the electronic relaxation components having amplitudes A$_1$, A$_2$ and A$_3$, fast oscillations ($l = 1$) attributed to coherent optical phonons and slow oscillations ($l = 2,3$) attributed to coherent acoustic phonons have been marked. The long wavelength oscillatory contribution in the signal could not be fitted using a single oscillator and a minimum of two oscillators are needed to fit the data as compared in Fig. S1(b) for three representative temperatures, each from the SC, SDW and the normal state of the systems where the data has been 5-point averaged for clarity.

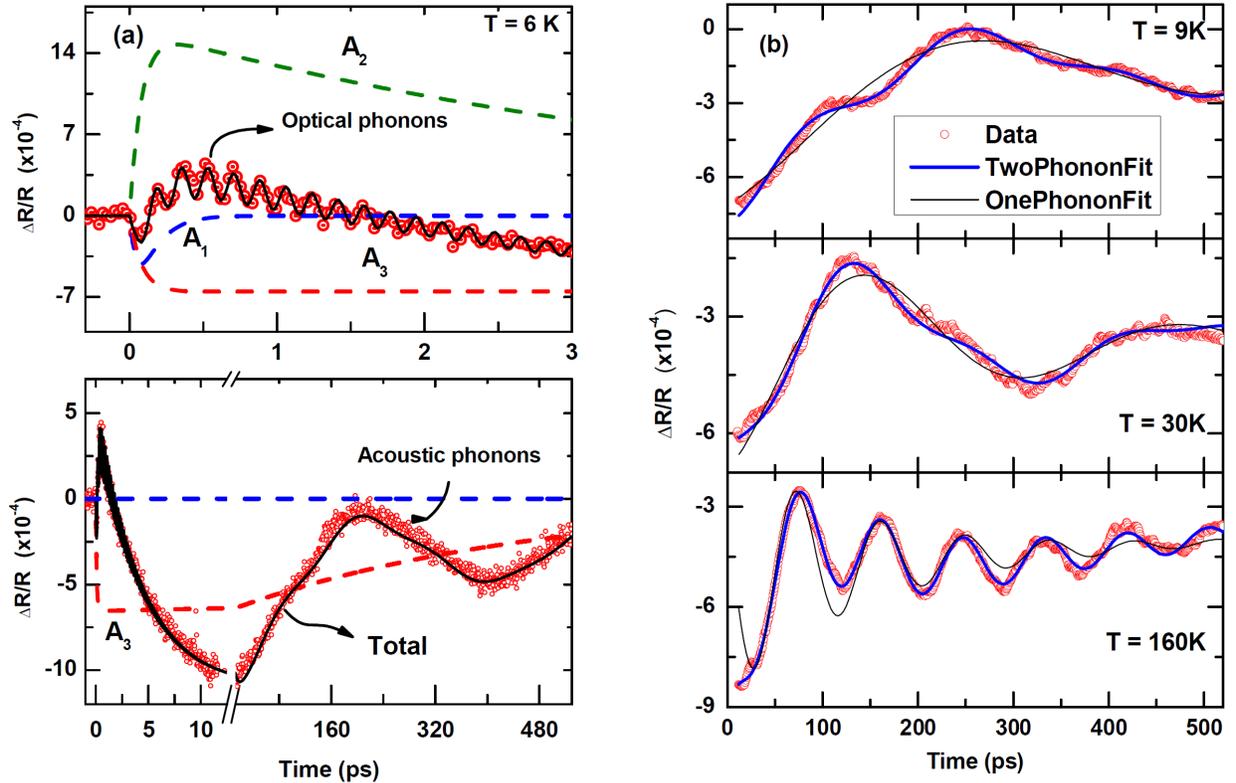

**Fig. S1.** Fitting procedure of the transient differential reflectivity data (open circles) using Eq. 1 at shorter and longer time-scales. (b) Comparison between two-phonon and one-phonon fit for the long wavelength coherent oscillations at three representative temperatures.